\documentclass[journal]{IEEEtran}
\usepackage{amsfonts}
\usepackage{amssymb}
\usepackage{amsmath}
\usepackage{mathrsfs}
\usepackage{verbatim}
\usepackage{subfigure}
\usepackage{balance}
\usepackage{booktabs}
\usepackage{fancybox}
\usepackage{bm}
\usepackage{extarrows}
\usepackage{algorithm}
\usepackage{algorithmic}
\usepackage{multirow}
\usepackage{array}
\usepackage{graphicx}
\usepackage{epstopdf} 
\newtheorem{theorem}{Theorem}
\newtheorem{lemma}{Lemma}

\begin{document}
%
%
\title{\LARGE{Dual User Selection for Security Enhancement \\ in Uplink Multiuser Systems}}

        \author{Hao~Deng,
        Hui-Ming~Wang,~\IEEEmembership{Senior Member,~IEEE,}
        Wenjie~Wang,~\IEEEmembership{Member,~IEEE,} \\
        and Moon Ho Lee,~\IEEEmembership{Life Senior Member,~IEEE}
\thanks{Manuscript received April 22, 2016; revised June 6, 2016; accepted June 17, 2016. Date of publication XXXXX, 2016; date of current version XXXXX, 2016. The work of H. Deng, H.-M. Wang and W. Wang was supported in part by the Foundation for the Author of National Excellent Doctoral Dissertation of China under Grant 201340, the National High-Tech Research and Development Program of China under Grant No.2015AA01A708, and the New Century Excellent Talents Support Fund of China under Grant NCET-13-0458. The work of
M. H. Lee was supported by MEST 2015 R1A2A1A05000977, NRF, Korea. The associate editor coordinating the review of this paper and approving it for publication was T. Q. Duong. \emph{(Corresponding author: Hui-Ming Wang.)}}
\thanks {H. Deng, H.-M. Wang and W. Wang are with the School
of Electronic and Information Engineering, Xi'an Jiaotong University,
Xi'an 710049, P. R. China, and H. Deng is also with the School of Physics and Electronics, Henan University, Kaifeng 475001,
P. R. China. (e-mail: gavind@163.com; xjbswhm@gmail.com; wjwang@xjtu.edu.cn).}
\thanks{M. H. Lee is with the Division of Electronics Engineering, Chonbuk National
University, Jeonju 561-756, South Korea (e-mail: moonho@jbnu.ac.kr).}
\thanks{Digital Object Identifier 10.1109/LCOMM.2016.XXXXX}

\vspace{-8pt}}

\maketitle
\begin{abstract}
 This letter proposes a novel dual user selection scheme for uplink transmission with multiple users, where a jamming user and a served user are jointly selected to improve the secrecy performance. Specifically, the jamming user transmits jamming signal with a certain rate, so that the base station (BS) can decode the jamming signal before detecting the secret information signal. By carefully selecting the jamming user and the served user, it makes the eavesdropper decode the jamming signal with a low probability meanwhile the BS can achieve a high receive signal-to-noise ratio (SNR). Therefore, the uplink transmission achieves the dual secrecy improvement by jamming and scheduling. It is shown that the proposed scheme can significantly improve the security of the uplink transmission.

\end{abstract}


\begin{IEEEkeywords}
 Physical layer security, ergodic secrecy rate, dual user selection.
\end{IEEEkeywords}

%
\IEEEpeerreviewmaketitle

\section{Introduction}
Physical layer security has emerged as a promising technique to improve the security of wireless communications \cite{S. Shetty}. Roughly speaking, a positive secrecy rate is achieved only if the main channel is better than the wiretap channel. However, it can not be always guaranteed due to channel fading. In order to make the wiretap channel ``worse'' than the main channel, the use of friendly jamming to selectively degrade the eavesdropper's channel has been considered in  \cite{Goel}-\cite{H. Deng}. When there are multiple users in the systems, user selection can make the BS-user pair facilitate a good channel conditions \cite{T. M. Hoang}-\cite{M. Pei}. It has been shown that the security can be significantly enhanced by friendly jamming or user selection.

In order to achieve both the benefit of friendly jamming and user selection, we propose a novel dual user selection scheme for uplink transmission without jamming signal coordination. Specifically, the user with the best channel condition to the BS will send jamming signals to create interference at the eavesdropper, while a selected served user will simultaneously transmit secret information signals to the BS. In the following, we call them jamming user and served user, respectively. By exploiting prior channel state information (CSI) of the main channels, the jamming user will transmit the jamming signal with a certain rate so that the BS can cancel out it by applying successive interference cancellation (SIC) \cite{Z. Ding}. In contrast, the eavesdropper would decode the jamming signal successfully with a low probability. Even when the eavesdropper decodes the jamming signal before detecting the information signal, the BS would achieve an higher average receive SNR than the eavesdropper by carefully selecting the served user. Therefore, secure communications can be definitely guaranteed.

\emph{Notations}: $\operatorname{C}_m^n$ denotes the combinatorial number $m$ over $n$, $\log(.)$ denotes base-$e$ logarithm, ${\operatorname{E}_{1}}\left( x \right)=\int_{x}^{\infty }{\frac{{{e}^{-t}}}{t}}dt$ is the zero-th incomplete gamma function, ${\operatorname{Li}_{2}}\left( x \right)=-\int_{0}^{x}{\frac{\log(1-t)}{t}}dt$ is the Euler dilogarithm function, and $\gamma =0.57721566\cdots$ is the Euler constant.

\section{System Model}
We consider an multiuser uplink transmission consisting of a base station (BS), $K$ users, and an eavesdropper. Each node is equipped with a single antenna. The quasi-stationary flat-fading channel coefficients from the $i$-th user to the BS and the eavesdropper, $h_i$ and $g_i$, are independent identically distributed (i.i.d.) complex Gaussian random variables with zero mean and unit variance, for $i=1,\cdots,K$. Without loss of generality, we consider that the users are ordered based on their channel quality, i.e., it holds $|h_1|^2 \le \cdots \le |h_K|^2$. The noises at the BS and the eavesdropper are assumed to be complex additive white Gaussian noise with zero mean and variance $\delta_n^2$.

In each time slot, two users are selected from the $K$ users, one jamming user and one served user. During the transmission, the jamming user sends jamming signals $z$ and the served user simultaneously transmits secret information signal $s$ with the same power $P/2$, where both $s$ and $z$ follow with $\mathcal{CN}(0,1)$. Suppose that the $m$-th user and the $n$-th user has been selected as the jamming user and the served user, respectively. Then the received signals at the BS and the eavesdropper can be expressed as
\begin{align}
y_b &= \sqrt{{P}/{2}}h_m z + \sqrt{{P}/{2}}h_n s + n_b, \\
y_e &= \sqrt{{P}/{2}}g_m z + \sqrt{{P}/{2}}g_n s + n_e,
\end{align}

The receive SNR for the BS to detect the jamming signal is given by
\begin{align}
\Gamma^{m,n}_{b,z} = \frac{|h_m|^2}{|h_n|^2+\frac{2}{\rho}}, \label{SNR_b_z}
\end{align}
where $\rho=\frac{P}{\delta_n^2}$ is the transmit SNR. The jamming user would send jamming signal with a rate $R_J$ satisfying $R_J \le \log(1+\Gamma^{m,n}_{b,z})$, so that the BS can cancel out the jamming signal before detecting the secure information signal. Hence, the achievable rate of the information signal at the BS is
\begin{align}
C_{b,s} = \log\left(1+\frac{\rho}{2}|h_n|^2\right).
\end{align}

The receive SNR for the eavesdropper to detect the jamming signal is given by
\begin{align}
\Gamma_{e,z}^{m,n} = \frac{|g_m|^2}{|g_n|^2+\frac{2}{\rho}}.
\end{align}

Similarly, we assume that SIC will be carried out at the eavesdropper. It will detect the jamming signal before decoding the secrecy information signal. If the rate of the jamming signal exceeds the capacity, i.e., $R_J > \log(1+\Gamma_{e,z}^{m,n})$, the eavesdropper can not decode the jamming signal successfully, and thus it treats the jamming signal as noise \cite{Z. Ding}. In contrast, if $R_J \le \log(1+\Gamma_{e,z}^{m,n})$, the jamming signal can be decoded successfully and thus cancelled out at the eavesdropper. Hence, the achievable rate at the eavesdropper is
\begin{align}
C_{e,s} =
\begin{aligned}
\begin{cases}
\log\left(1+\frac{\rho}{2}|g_n|^2\right),&R_J \le \log(1+\Gamma_{e,z}^{m,n}), \\
\log\left(1+\frac{|g_n|^2}{|g_m|^2+\frac{2}{\rho}}\right),& R_J > \log(1+\Gamma_{e,z}^{m,n}).
\end{cases}
\end{aligned}
\end{align}

Typically, for a passive eavesdropper, its CSI can not be available at the BS. Therefore, we use ergodic secrecy rate (ESR) as the performance metric, which is defined as \cite{Y. Liang}
\begin{align}
R_s =\Big\{\mathbb{E} \left[C_{b,s}\right] -\mathbb{E}\left[C_{e,s}\right]\Big\}^+,  \label{ESR}
\end{align}
where $\{x\}^+ = \max(0,x)$.

Obviously, the selections of jamming user and served user would impact greatly on the secrecy performance of the proposed scheme. In the following, we will give the details about how to select the jamming user and the served user.

\section{Jamming User and Served User Selection}
\subsection{Jamming User Selection}
We first focus on the selection of jamming user. Since when $R_J \le \log(1+\Gamma_{e,z}^{m,n})$, the eavesdropper can decode the jamming signal before detecting the secret information signal. This implies that the jamming user should transmit the jamming signal with a rate as high as possible to defeat the potential eavesdropping. Therefore, the jamming user should transmit the jamming signal with a rate up to the capacity $\log(1+\Gamma_{b,z}^{m,n})$. Observing Eqn. \eqref{SNR_b_z}, we can know that when the jamming user and the served users are the pair who satisfies that $|h|_m^2$ is much larger than $|h|_n^2$, it would achieves a large value of $\Gamma_{b,z}^{m,n}$. This makes the eavesdropper achieve a low probability to decode the jamming signal successfully. Hence, it is reasonable to select the user with the best channel condition as the jamming user. That is, the $K$-th user operates as a jamming user and will transmit the jamming signal with a rate $\log(1+\Gamma_{b,z}^{K,n})$. Now we focus on the selection of served user. Since the secrecy rate is the difference between the rates achieved at the BS and the eavesdropper, it requires that the served user should have a good channel condition. However, selecting a user with good channel channel condition as the served user would reduce the jamming rate. Therefore, there is a tradeoff between the jamming rate and the achievable rate at the BS. In the following, we will determinate the served user by finding the $n$-th user who achieves the maximum ESR.
\subsection{Auxiliary Results}
In this subsection, we provide the following lemma that will be used in the ESR analysis.
\begin{lemma}
Define $T = \frac{|h_K|^2}{|h_n|^2+\frac{2}{\rho}}$, its cumulative distribution function (CDF) is
\begin{align}
F_T(t) =
\begin{cases}
\sum_{i=0}^{K-n}\sum_{j=0}^{n-1} \frac{\Xi_{ij} e^{-\frac{2t i}{\rho }}}{i(t-1)+K-n+1+j},\hspace{1.1cm} {t \ge 1}, \\
\sum_{i=0}^{K-n}\sum_{j=0}^{n-1}\Xi_{ij}\frac{e^{-\frac{ 2t i}{\rho}}-e^{-\frac{2 t (K-n+1+j)}{\rho(1-t)}}}{i(t-1)+K-n+1+j},{0 \le t < 1},
\end{cases} \nonumber
\end{align}
where
\begin{align} \label{Xi_ij}
\Xi_{ij}\triangleq \begin{cases}
(-1)^{i} n\operatorname{C}_K^n  \operatorname{C}_{K-n}^i, &j=0,\\
(-1)^{i+j} n\operatorname{C}_K^n  \operatorname{C}_{K-n}^i \frac{(n-1)\cdots(n-j)}{j!},&j \ge 1.
\end{cases}
\end{align}
\end{lemma}
\begin{IEEEproof}
Please see Appendix A.
\end{IEEEproof}

\subsection{ESR and Served User Selection}
In this section, we give the served user selection scheme to achieve the maximal ESR. The exact ESR of the proposed scheme is given by the following theorem.
\begin{theorem}
When the $n$-th user is selected as the served user and the $K$-th user operates as a jamming user, the achievable ESR can be expressed as
\begin{align}
&R_s = \Bigg\{\sum_{i=1}^K(-1)^{i+1}\operatorname{C}_K^i e^{\frac{2i}{\rho}}\operatorname{E}_1\left(\frac{2i}{\rho}\right)
-\sum_{i=n}^{K-1}\sum_{j=0}^{i}(-1)^{j}\operatorname{C}_K^i \operatorname{C}_i^j
\nonumber \\
&e^{\frac{2(K+j-i)}{\rho}}\operatorname{E}_1\left(\frac{2(K+j-i)}{\rho}\right)-1+\frac{2}{\rho}e^{\frac{2}{\rho}}
\operatorname{E}_1\left(\frac{2}{\rho}\right)- e^{\frac{2}{\rho}}\Psi
\Bigg\}^+, \nonumber
\end{align}
where $\Theta(u) \triangleq\frac{e^\frac{2\eta}{\rho }\operatorname{E}_{1}\big(\frac{2\eta}{\rho}\big)+1-\log(u+1)}{(u+1)^2}-\frac{\frac{2}{\rho}e^\frac{2\eta}{\rho }\operatorname{E}_{1}\big(\frac{2\eta}{\rho}\big)}{u (u+1)}$, $\eta = \frac{u+1}{u}$, and $\Psi=\int_0^{\infty}\Theta(u)F_T(u)du $.
\end{theorem}
\begin{IEEEproof}
Please see Appendix B.
\end{IEEEproof}

 Although Theorem 1 does not provide a closed form expression for $R_s$, it can be easily obtained by numerical computation. Using Theorem 1, the index of the served user can be determined by one dimension search.

 Next, we investigate the asymptotic behavior of the ESR in the high SNR regime. An approximation of the result in Theorem 1 is given in the following Corollary.

\emph{Corollary 1:} In the high SNR regime $(\rho \rightarrow \infty)$, a closed-form expression of the ESR is given as
\begin{align}
R_s^{Hi} &\thickapprox \Big\{\frac{\log\frac{\rho}{2}-1-\gamma}{2}+\varpi-\sum_{i=1}^{K-n}\sum_{j=0}^{n-1}  \frac{\Xi_{ij}}{i}\Upsilon_{ij}\Big\}^+,
\end{align}
where
\begin{align}
\Upsilon_{ij}=
\begin{cases}
  \frac{ \log\frac{\rho}{2}+1-\gamma}{8} + \frac{\log2}{4} - \frac{3}{8},&\xi_{ij} =1,\\
  \nu_{ij}+\frac{2\big(\operatorname{Li}_2(\frac{\xi_{ij}-1}{\xi_{ij}})
  -\operatorname{Li}_2(\frac{\xi_{ij}-1}{2\xi_{ij}})\big)-\operatorname{Li}_2(-{\xi_{ij}})+\zeta_1}{(1-\xi_{ij})^2},&\xi_{ij} <1, \\
  \nu_{ij}+\frac{2\big(\operatorname{Li}_2(\frac{\xi_{ij}-1}{\xi_{ij}})
  -\operatorname{Li}_2(\frac{\xi_{ij}-1}{2\xi_{ij}})\big)-\operatorname{Li}_2(-{\xi_{ij}})+\zeta_2}{(1-\xi_{ij})^2},&\xi_{ij} >1.
\end{cases} \nonumber
\end{align}
$\varpi$, $\nu_{ij}$, $\xi_{ij}$, $\zeta_1$ and $\zeta_2$ are defined in Appendix C.
\begin{IEEEproof}
Please see Appendix C.
\end{IEEEproof}

In order to highlight the secrecy improvement, let us compare the ESR of our scheme to that of the TDMA-like scheme \cite{M. Pei}, where the user with the best channel condition transmits the secure information signal with power $P$ to the BS in each time slot. Similarly, the ESR of the TDMA-like one in the high SNR regime is approximated as
\begin{align}
R_{s,TD}^{Hi} \approx \Big\{\sum_{i=1}^K(-1)^{i+1}\operatorname{C}_K^i \log i\Big\}^+. \label{R_s_TD}
\end{align}

As can be seen from \eqref{R_s_TD}, increasing transmit power does not improve the ESR of the TDMA-like scheme. In contrast, the ESR of our scheme scales with $\log{\frac{\rho}{2}}$ in the high SNR regime. Since the served user is not the one with the best channel condition, our scheme achieves a less multiuser gain than the TDMA-like scheme. However, the use of jamming helps the BS to benefit from high transmit power, thus it significantly improve the secrecy performance.

\section{Simlulations}
\begin{figure}[t]
\begin{center}
\includegraphics[width=3 in]{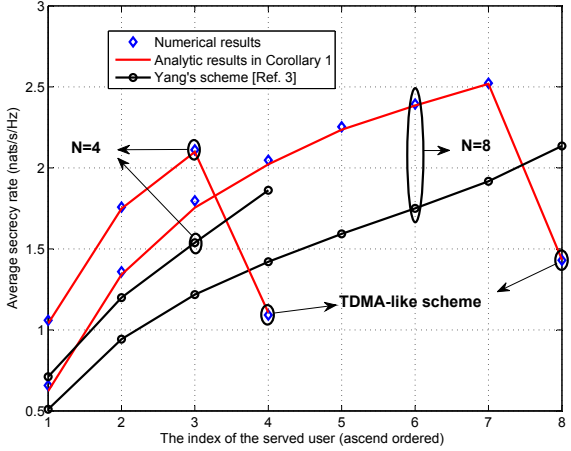}
\end{center}
\vspace{-0.3cm}
\caption{ESR vs. the index of the served user.}
\end{figure}

We plot the analytic result in Corollary 1 and the numerical result in Fig. 1. The numerical results are obtained by performing Monte Carlo experiments consisting of 10000 independent trials. We set $\rho = 20 \ dB$. This figure illustrates the achievable ESR vs. the index of the served user. Note that, when the served user is the $K$-th user, no user is employed to transmit jamming signal, it is the TDMA-like scheme proposed in \cite{M. Pei}. For comparison, the secrecy rate of Yang's scheme proposed in \cite{Yang} is also plotted, where all the users except the served user operate as helpers and the power budget of each node in the transmission is assumed to be $P$. Hence, Yang's scheme consumes more power than our proposed scheme.

Fig. 1 shows that the analytic results agree well with the numerical results even when $\rho = 20 \ dB$. As can be seen, the optimal user to be served is the $3$-th user when $N=4$, while it is the $7$-th user when $N = 8$. It is shown that when the best user is selected to communicate with, the ESR of our scheme achieves a much higher secrecy rate than that of the TDMA-like scheme and Yang's scheme. This is because the TDMA-like scheme gains no benefit from jamming while Yang's scheme requires two phases to coordinate the jamming signals in each date transmission round.

\section{Conclusions}
This letter proposed a novel dual user selection scheme to enhance the security of uplink transmission.  Analytical results of ESR for the proposed scheme have been developed. It has been recognized that the proposed scheme can benefit from multiuser diversity gain and friendly jamming, which significantly enhances the security.

\appendices
\section{}
Let $X = |h_n|^2$ and $Y=|h_K|^2$. The joint probability density functions of the two order statistics, $X$ and $Y$, is given as \cite{David}
\begin{align}
&f_{X,Y}(x,y) \nonumber \\
& = \operatorname{C}_K^{n-1}{(1-e^{-x})^{n-1}}(e^{-x}-e^{-y})^{K-n-1}e^{-x}e^{-y}, \ \ x<y. \nonumber
\end{align}

Define $T = \frac{|h_K|^2}{|h_n|^2+\frac{2}{\rho}}$. The CDF of $T$ can be calculated as
\begin{align}
 F_T(t) &= \mathbb{P}\left\{\frac{y}{x+2/\rho} \le t\right\} \nonumber \\
&=
\begin{aligned}
\begin{cases}
\operatorname{C}_K^{n-1}\int_0^{\infty}(1-e^{-x})^{n-1}e^{-x}dx \\
\int_x^{xt+\frac{2t}{\rho}}(e^{-x}-e^{-y})^{K-n-1}e^{-y}dy,& t \ge 1, \\
\operatorname{C}_K^{n-1}\int_0^{\frac{2t}{\rho(1-t)}}(1-e^{-x})^{n-1}e^{-x}dx \\
\int_x^{xt+\frac{2t}{\rho}}(e^{-x}-e^{-y})^{K-n-1}e^{-y}dy,& 0 \le t < 1.
\end{cases}
\end{aligned} \nonumber
\end{align}

When $t \ge 1$, we further have
\begin{align}
F_T(t)& = \frac{\operatorname{C}_K^{n-1}}{K-n}\int_0^{\infty}(1-e^{-x})^{n-1}e^{-x}(e^{-x}-e^{-(xt+\frac{2t}{\rho})})^{K-n}dx \nonumber \\
& = \sum_{i=0}^{K-n}\Xi_{i0}\int_0^{\infty}(1-e^{-x})^{n-1}e^{-\frac{2 t i}{\rho}-\left(i(t-1)+K-n+1\right)x}dx \nonumber \\
& = \sum_{i=0}^{K-n}\sum_{j=0}^{n-1}\Xi_{ij}e^{-\frac{2t i}{\rho }}\int_0^{\infty} e^{-(i(t-1)+K-n+1+j)x}dx \nonumber \\
& = \sum_{i=0}^{K-n}\sum_{j=0}^{n-1} \frac{\Xi_{ij} e^{-\frac{2 t i}{\rho}}}{i(t-1)+K-n+1+j}.
\end{align}
where ${\Xi_{ij}}$ is defined in \eqref{Xi_ij}. Similarly, we can proof the case $0<t<1$.
\section{}

Let $X= |h_n|^2$. The CDF of the $n$-th order statistics, $|h_n|^2$, is given by \cite{David}
\begin{align}
F_X(x)= \sum_{i=n}^K\operatorname{C}_K^i(1-e^{-x})^i (e^{-x})^{K-i}.
\end{align}

Then, the expectation of $C_{b,s}$ can be calculated as
\begin{align}
&\mathbb{E} \left[C_{b,s}\right]=\int_0^\infty\frac{\frac{\rho}{2}(1-F_X(x))}{1+\frac{\rho}{2}x}dx=\sum_{i=1}^K(-1)^{i+1}\operatorname{C}_K^i e^{\frac{2i}{\rho}}\operatorname{E}_1\left(\frac{2i}{\rho}\right)
\nonumber \\
&-\sum_{i=n}^{K-1}\sum_{j=0}^{i}(-1)^{j}\operatorname{C}_K^i \operatorname{C}_i^j e^{\frac{2(K+j-i)}{\rho}}\operatorname{E}_1\left(\frac{2(K+j-i)}{\rho}\right). \label{Rb}
\end{align}

Next, we calculate the expectation of $C_{e,s}$. As we have discussed in Section III-A, the $K$-th user operates as a jamming user and transmit the jamming signal with a rate $\log(1+\Gamma_{b,z}^{K,n})$. The achievable rate at the eavesdropper can be further given as
\begin{align}
C_{e,s} =
\begin{aligned}
\begin{cases}
\log\left(1+\frac{\rho}{2}|g_n|^2\right),&\Gamma_{b,z}^{K,n} \le \Gamma_{e,z}^{K,n}, \\
\log\left(1+\frac{|g_n|^2}{|g_K|^2+\frac{2}{\rho}}\right),& \Gamma_{b,z}^{K,n} > \Gamma_{e,z}^{K,n}.
\end{cases}
\end{aligned}
\end{align}

Let $T = \Gamma_{b,z}^{K,n} = \frac{|h_K|^2}{|h_n|^2+\frac{2}{\rho}}$, $Y = |g_K|^2$ and $Z=|g_n|^2$. The expectation of $C_{e,s}$ is given as
\begin{align}
&\mathbb{E} \left[C_{e,s}\right] = \int_0^{\frac{y}{z+\frac{2}{\rho}}}dF_T(t)\int_0^{\infty}\int_0^{\infty}\log\Big(1+\frac{\rho}{2}z\Big)e^{-z-y}dzdy \nonumber \\
&+\int_{\frac{y}{z+\frac{2}{\rho}}}^0 dF_T(t)\int_0^{\infty}\int_0^{\infty}\log\frac{z+y+\frac{2}{\rho}}{y+\frac{2}{\rho}}e^{-z-y}dzdy  \nonumber \\
&\ \ \ \ \ \ \ \ \ =\int_0^{\infty}\int_0^{\infty}\log\frac{z+y+\frac{2}{\rho}}{y+\frac{2}{\rho}}e^{-z-y}dzdy\nonumber \\
&+\int_0^{\infty}\int_0^{\infty} F_T\Big(\frac{y}{z+\frac{2}{\rho}}\Big)\log\frac{(1+\frac{\rho}{2}z)
(y+\frac{2}{\rho})}{z+y+\frac{2}{\rho}}e^{-z-y}dzdy. \label{E_Ce}
\end{align}

Let $R_{e1}$ and $R_{e2}$ denote the first term and the second term of the right-hand-side (RHS) in \eqref{E_Ce}, respectively. The expression of $R_{e1}$ can be easily given as
\begin{align}
R_{e1} = 1-\frac{2}{\rho}e^{\frac{2}{\rho}}\operatorname{E}_1\left(\frac{2}{\rho}\right). \label{Re1}
\end{align}

We further define $u=\frac{y}{z+\frac{2}{\rho}}$ and $v=y$. The Jacobian matrix is $J = \left|
\begin{array}{lll}
\frac{dz}{du} & \frac{dz}{dv}\\
\frac{dy}{du}& \frac{dy}{dv}
\end{array}\right| =\left|
\begin{array}{lll}
-\frac{v}{u^2} & \frac{1}{u}\\
0& 1
\end{array}\right|$, and its determinant is $|J| = -\frac{v}{u^2}$. Therefore, $R_{e2}$ can be calculated as
\begin{align}
R_{e2} &= e^{\frac{2}{\rho}}\int_0^{\infty}\int_0^{\infty}\log\frac{1+\frac{\rho}{2}v}{1+u}\frac{vF_T(u)}{u^2}e^{-\frac{u+1}{u}v}dudv \nonumber \\
&=e^{\frac{2}{\rho}}\int_0^{\infty}\Big(\frac{e^\frac{2(u+1)}{\rho u}\operatorname{E}_{1}\big(\frac{2(u+1)}{\rho u}\big)+1-\log(u+1)}{(u+1)^2}\nonumber \\
&\ \ \ -\frac{\frac{2}{\rho}e^\frac{2(u+1)}{\rho u}\operatorname{E}_{1}\big(\frac{2(u+1)}{\rho u}\big)}{ u (u+1)}\Big)F_T(u)du. \label{Re2}
\end{align}

Substituting the results in \eqref{Rb},  \eqref{Re1}, and \eqref{Re2} into \eqref{ESR} yields the result in Theorem 1. The proof is completed.

\section{}
Using the result $e^{\frac{1}{x}}\operatorname{E}_1\left(\frac{1}{x}\right) \thickapprox \log x - \gamma$ as $x \rightarrow \infty$ \cite{H. Deng}, the expectation of $C_{b,s}$ at high SNR can be approximated as
\begin{align}
\mathbb{E} \left[C_{b,s}\right] \thickapprox &\log\frac{\rho}{2} - \gamma +\varpi. \label{Cb}
\end{align}
where $\varpi\triangleq
-\sum_{i=n}^{K-1}\sum_{j=0}^{i}(-1)^{j+1}\operatorname{C}_K^i \operatorname{C}_i^j \log(K+j-i)+\sum_{i=1}^K(-1)^{i}\operatorname{C}_K^i \log i $.
As $\rho \rightarrow \infty$, it always holds $T = \frac{|h_K|^2}{|h_n|^2+\frac{2}{\rho}} \rightarrow \frac{|h_K|^2}{|h_n|^2} \ge 1$, and thus it holds $F_T(t) = 0$ when $0 \le t < 1$. Therefore, the approximation of $F_T(t)$ at high SNR, denoted by $F_T^{Hi}(t)$, can be given as
 \begin{align}
F_T^{Hi}(t) =
\begin{cases}
\sum_{i=0}^{K-n}\sum_{j=0}^{n-1} \frac{\Xi_{ij}}{i(t-1)+K-n+1+j},&{t \ge 1}, \\
0,&{0 \le t < 1}.
\end{cases} \nonumber
\end{align}

Then, the expectation of $C_{e,s}$ in the high SNR regime can be approximated as
\begin{align}
\mathbb{E} \left[C_{e,s}\right]
\thickapprox & 1+ \int_0^{\infty}\frac{\log\frac{\rho}{2}+1-\gamma+\log\frac{u}{(u+1)^2}}{(u+1)^2}F_T^{Hi}(u)du \nonumber \\
=&\frac{\log\frac{\rho}{2}+1-\gamma}{2}+\sum_{i=1}^{K-n}\sum_{j=0}^{n-1}  \frac{\Xi_{ij}}{i}\Upsilon_{ij}, \label{Ce_1}
\end{align}
where
\begin{align}
\Upsilon_{ij} &\triangleq\int_1^{\infty}\frac{ \log\frac{\rho}{2}+1-\gamma+\log\frac{u}{(u+1)^2}}{(u+1)^2(u+\xi_{ij})}du,
\end{align}
\begin{align}
\xi_{ij} &\triangleq \frac{K-n+1+j}{i}-1.
\end{align}

To obtain a closed-form expression for $\Upsilon_{ij}$, we should consider two cases, $\xi_{ij}=1$ and $\xi_{ij} \neq 1$. When $\xi_{ij}=1$, we have
\begin{align}
\Upsilon_{ij} &= \int_1^{\infty}\frac{ \log\frac{\rho}{2}+1-\gamma+\log\frac{u}{(u+1)^2}}{(u+1)^3}du \nonumber \\
&= \frac{ \log\frac{\rho}{2}+1-\gamma}{8} + \frac{\log2}{4} - \frac{3}{8}.
\end{align}

When $\xi_{ij} \neq 1$, we have
\begin{align}
\Upsilon_{ij} &= \int_1^{\infty}\frac{ \log\frac{\rho}{2}+1-\gamma+\log\frac{u}{(u+1)^2}}{(u+1)^2(u+\xi_{ij})}du \nonumber \\
&= \int_1^{\infty}\frac{\log\frac{\rho}{2}+1-\gamma}{(u+1)^2(u+\xi_{ij})}du+\int_1^{\infty}\frac{\log\frac{(u+1)^2}{u}}{(1-\xi_{ij})(u+1)^2}du \nonumber \\
&\ \ \ +\frac{1}{1-\xi_{ij}}\int_1^{\infty}\frac{\log\frac{u}{(u+1)^2}}{(u+1)(u+\xi_{ij})}du \nonumber \\
&=\big(\log\frac{\rho}{2}+1-\gamma\big)\frac{\xi_{ij}-1+2\log\frac{2}{1+\xi_{ij}}}{2(1-\xi_{ij})^2}+\frac{1}{1-\xi_{ij}} \nonumber \\
&\ \ \ +\frac{1}{\xi_{ij}(1-\xi_{ij})}\int_0^{1}\frac{\log\frac{v}{(v+1)^2}}{(v+1)(v+1/\xi_{ij})}dv \nonumber \\
&=\nu_{ij}+\mu_{ij},
\end{align}
where $\nu_{ij}\triangleq\big(\log\frac{\rho}{2}+1-\gamma\big)\frac{\xi_{ij}-1+2\log\frac{2}{1+\xi_{ij}}}{2(1-\xi_{ij})^2}+\frac{1}{1-\xi_{ij}} -\frac{\pi^2+12\log^2 2}{12(1-\xi_{ij})^2}$ and
\begin{align}
\mu_{ij}=
\begin{cases}
\frac{2\big(\operatorname{Li}_2(\frac{\xi_{ij}-1}{\xi_{ij}})-\operatorname{Li}_2(\frac{\xi_{ij}-1}{2\xi_{ij}})\big)-\operatorname{Li}_2(-{\xi_{ij}})+\zeta_1}{(1-\xi_{ij})^2},&\xi_{ij} <1, \\
\frac{2\big(\operatorname{Li}_2(\frac{\xi_{ij}-1}{\xi_{ij}})-\operatorname{Li}_2(\frac{\xi_{ij}-1}{2\xi_{ij}})\big)-\operatorname{Li}_2(-{\xi_{ij}})+\zeta_2}{(1-\xi_{ij})^2},&\xi_{ij}>1,
\end{cases} \nonumber
\end{align}
where $\zeta_1\triangleq2\log 2 \log\frac{\xi_{ij}+1}{\xi_{ij}}-{\log^2 2}$ and $\zeta_2\triangleq 2\log 2 \log \frac{\xi_{ij}+1}{\xi_{ij}-1}+{\log^2\frac{\xi_{ij}-1}{\xi_{ij}}-\log^2\frac{\xi_{ij}-1}{2\xi_{ij}}}$. Combining \eqref{Cb} and \eqref{Ce_1}, the result in Corollary 1 is obtained.

\end{document}